\documentclass[aps,twocolumn,superscriptaddress,footinbib,floatfix,showpacs]{revtex4} 
\usepackage{graphicx} 
\usepackage{float}
\include{test.tex}
  
\begin{document} 
\title{Semiclassics of the Chaotic Quantum-Classical Transition}  
      \vbox to 0pt{\vss 
                    \hbox to 0pt{\hskip-40pt\rm LA-UR-03-9308\hss} 
                   \vskip 25pt} \preprint{LA-UR-03-9308}
                   \author{Benjamin D. Greenbaum}
                   \affiliation{Department of Physics, Columbia
                   University, New York, New York 10027}
                   \author{Salman Habib} \affiliation{T-8, Theoretical
                   Division, Los Alamos National Laboratory, Los
                   Alamos, NM 87545} \author{Kosuke Shizume}
                   \affiliation{Institute of Library and Information
                   Science, University of Tsukuba, 1-2 Kasuga,
                   Tsukuba, Ibaraki 305-8550, Japan} \author{Bala
                   Sundaram} \affiliation{Department of Physics,
                   University of Massachusetts, Boston, MA 02125}
                   \begin{abstract}
  
                     {We elucidate the basic physical mechanisms
                       responsible for the quantum-classical
                       transition in one-dimensional, bounded chaotic
                       systems subject to unconditioned environmental
                       interactions.  We show that such a transition
                       occurs due to the dual role of noise in
                       regularizing the semiclassical Wigner function
                       and averaging over fine structures in classical
                       phase space.  The results are interpreted in
                       the novel context of applying recent advances
                       in the theory of measurement and open systems
                       to the semiclassical quantum regime.  We use
                       these methods to show how a local semiclassical
                       picture is stabilized and can then be
                       approximated by a classical distribution at
                       later times.  The general results are
                       demonstrated explicitly via high-resolution
                       numerical simulations of the quantum master
                       equation for a chaotic Duffing oscillator.}

\end{abstract} 

\pacs{05.45.Mt,03.65.Sq,03.65.Bz,65.50.+m} 

\maketitle

\section{Introduction}

Ever since the birth of quantum physics, the boundary between quantum
and classical descriptions of nature has been the cause of much
controversy and debate.  Although few people now believe in the
required existence of a ``large'' classical world in which quantum
mechanics is somehow embedded, even for those that accept the primacy
of a full quantum description, the identification of the actual
physical processes that allow a quantum dynamical system to be
approximated -- in some limit -- by a classical dynamical system often
remains less than clear-cut.

Initially, quantum-classical correspondence was phrased in the context
of understanding how the fundamental ``subatomic'' laws of quantum
physics could possibly be compatible with a ``macroscopic'' world
which, to a very good degree of approximation, evolves according to
classical Hamiltonian dynamics and lacks (classically) bizarre quantum
characteristics such as interference and entanglement~\cite{Lan}.
This view was famously, if somewhat vaguely, canonized in Bohr's
Correspondence Principle.  The phrase is typically invoked to mean one
of three related, but not identical, subjects: the existence of a
formal analogy between certain preferred classical dynamical variables
and quantum observables; the limit of large quantum numbers, large
action or small $\hbar$, possibly in some combination; or the extent
to which classical and quantum dynamical evolutions agree, in the
spirit of Ehrenfest's theorem and semiclassical dynamics.

The last two interpretations, which are the principal foci of this
paper, often overlap with one another but are not identical.  As an
example, the position and momentum expectation values of a quantum
harmonic oscillator evolve exactly according to the classical
Liouville equation, and, given an initial distribution acceptable both
classically and quantum-mechanically, the two theories give identical
results. However, when comparing a quantum energy eigenstate of the
oscillator to a classical orbit at the same energy, ad hoc reasoning
must be utilized to eliminate rapid quantum oscillations about the
classical values, an example of the singular nature of the
$\hbar\rightarrow 0$ limit.  That said, interference elimination and
dynamical agreement are often related insofar as decreasing the size
of $\hbar$ will usually have the effect of altering the scale of
quantum interference while simultaneously improving the timescale of
agreement of classical and quantum expectation values when they are
not already identical (e.g., the trivial linear case above).

It has long been recognized that the problems with attaining the
classical limit are compounded for nonlinear
systems~\cite{BornEin}. Theoretical analysis and experimental
observation of chaotic systems over the past forty years has made it
clear that classical chaos is a real-world phenomenon that quantum
theory should reproduce to within experimental accuracy.  Under a
unitary quantum evolution, however, any nonlinear dynamical system
will eventually fail the conditions of Ehrenfest's theorem. Quantum
expectation values cannot follow classical predictions at long times
as quantum mechanics does not respect the symplectic dynamical
symmetry of classical mechanics~\cite{HabibDragt, hjmrss}.  The dynamics
of closed bounded quantum systems are also quasi-periodic; such a
system can never be chaotic for any non-zero value of $\hbar$.

A chaotic classical phase space evolution generates structures at
infinitesimally small scales, whereas, due to interference effects,
the corresponding quantum evolution does not possess a notion of local
phase space structures. The net effect is a short-time disagreement
between semiclassical and classical evolutions, followed by a failure
of the semiclassical approximation itself at longer, but still finite
timescales~\cite{BerryZas, heller}.  This prompted some early
investigators in the field to wonder if quantum mechanics had to be
modified in order to produce chaos~\cite{FordEtc}.  As a consequence
of these obstructions, chaotic systems have emerged as a testing
ground for whether or not quantum-classical correspondence is truly a
valid concept, and, if so, how it should be properly phrased and
addressed.

A parallel set of experimental developments, particularly in the last
twenty years, have also strongly suggested the need for a more refined
view of the quantum-classical transition (QCT).  The border between
the macro-world of classical mechanics and the micro-world of quantum
physics has been blurred by technological, observational, and
theoretical progress.  Precision measurements in nanomechanics, atomic
and molecular optics, and quantum information processing and
communication have probed mesoscopic regimes, necessitating a careful
analysis of the relative merits of using a classical or quantum
description since the systems studied are neither ``very large'' nor
``very small''. In a quite different realm, recent observations of the
cosmic microwave background and the large-scale distribution of
galaxies have strongly supported the notion that primordial quantum
fluctuations seed the formation of large scale structures in the
Universe~\cite{Wmap}, demonstrating that crude criteria of
``microscopic'' vs. ``macroscopic'' are no longer sufficient as an
underlying basis for a serious study of the QCT. Understanding the
physical mechanisms which define when a system behaves classically is
now a practical issue.

A consensus is forming that spanning the gap between the above
problems and the correspondence principle requires a robust
understanding of open quantum systems and quantum
measurement~\cite{Meas}.  Any experimentally relevant system is, by
definition, a {\em measured} system which interacts with its
environment, if only through a meter.  A quantum measurement differs
from a classical one in at least two regards: (i) the intrinsic
barrier imposed by the uncertainty principle on the precision of phase
space information a meter can extract and (ii) the more severe manner
in which the subsystem becomes entangled with its environment. Due to
this entanglement, quantum measurement is generically associated with
an irreducible disturbance on the observed system (quantum
``backaction''). The desired measurement process must yield a limited
amount of information in a finite time in order to yield dynamical
information without strongly influencing the dynamics. Hence, simple
projective (von Neumann) measurements are clearly not appropriate
because they yield complete information instantaneously via state
projection. But this fundamental notion of measurement can be easily
extended to devise schemes that extract information
continuously~\cite{genmeas}.

The basic idea is to have the system of interest interact weakly with
another (e.g., atom interacting with an electromagnetic field) and
make projective measurements on the auxiliary system (e.g., photon
counting). Because the interaction is small, the state of the
auxiliary system gathers little information regarding the system of
interest, and this system, in turn, is only perturbed slightly by the
measurement backaction. Only a small component of the information
gathered by the projective measurement of the auxiliary system relates
to the system of interest, and a continuous limit of the measurement
process can be taken.  One then studies the master equation for the
evolution of the subsystem density matrix conditioned on its
measurement record.  The master equation can be further ``unraveled''
into nonlinear stochastic trajectories for a pure state, the so-called
quantum trajectories~\cite{WisePhd}. An average over the pure states
gives back the original density matrix.  Unlike in the classical case,
where the analogous situation refers to a weighted ensemble of
phase-space points uniquely determined by the probability
distribution, a mixed-state density matrix does not have a unique
decomposition in terms of state vectors.

It is essential to distinguish between closed evolution, where the
system state evolves without any coupling to the external world, {\em
  unconditioned open} evolution, where the system evolves coupled to
an external environment but where no information regarding the system
is extracted from the environment, and {\em conditioned open}
evolution where such information {\em is} extracted.  What we call the
{\em strong} form of the QCT describes how a local trajectory level
picture arises from a conditioned evolution.  However, in many
situations, only a statistical description is possible even
classically, and here we will demand only the agreement of quantum and
classical distributions and the associated dynamical averages. This
defines the unconditioned {\em weak} form of the QCT which is the
focus of the present paper (for a review see Ref.~\cite{hekrev}).

While the specific nature of the subsystem-environment interaction
depends on the subsystem studied, the actual process of information
extraction, and unavoidable coupling to other environmental channels,
there do exist simple, yet physically significant, general cases. The
systems studied in this paper can be interpreted as undergoing a
continuous position measurement~\cite{CM}, where either the results of
measurement are not recorded, or all of the measurements in an
ensemble are averaged over to erase the information regarding specific
measurements. Nevertheless, the entanglement between the position
measuring readout and the subsystem still produces a quantum
backaction in momentum.  The form of this open system interaction,
which falls into the class of Lindblad superoperators, rigidly
separates the subsystem and its environment~\cite{Math}.

Although a classically chaotic system cannot approximate a closed
quantum system via the traditional $\hbar\rightarrow 0$ route, there
is good numerical evidence -- at least for some systems -- for the
weak form of the QCT.  Numerical studies of the Duffing oscillator and
other systems have shown that expectation values of a quantum
system subject to an unconditioned continuous position measurement
will come into agreement with the expectation values of an (equivalent)
open classical system, and that the quantum phase space will come to
capture certain classical phase space features~\cite{SalKo}. In the
case of the strong form of the QCT, studies have demonstrated the
existence of nonzero Lyapunov exponents for conditioned systems, as
well as inequalities which clearly delineate when the classical
trajectory interpretation is valid in the conditioned
case~\cite{Tan,SSJ}.
 
An important distinction between the weak and strong forms of the QCT
must be made.  In the conditioned case, the master equation actively
{\it localizes} the wavefunction about its expectation value, allowing
{\it trajectory} level agreement between measured classical and
quantum systems.  However, in the unconditioned case, the inequalities
governing the strong classical limit need not be satisfied and
localization need not occur.  The problem of understanding how
classical and quantum systems begin to look like one another in a
generic open system, even without the advantage of conditioning, has
remained open.

As a final point, we note that while the strong form of the QCT must
hold for all dynamical systems with a classical counterpart, it is not
that the weak QCT must also do so.  The quantum delta-kicked rotor
provides a particular example of the failure of the weak
QCT~\cite{bhjs}. The general problem of knowing in advance what
governs this behavior is not yet resolved, although the work in this
paper suggests that (effective) compactness of the accessible phase
space plays an important role.  Moreover, the violation of the
conditions necessary to establish the strong form of the QCT need not
prevent the existence of a weak QCT. Since the strong form of the QCT
requires treating the localized limit, a cumulant expansion for the
distribution function immediately suggests itself~\cite{Tan}, whereas,
for the more nonlocal issues relevant to the weak form of the QCT, a
semiclassical analysis turns out to be natural, as will be
demonstrated here.

In this paper we investigate the physical mechanisms responsible for
the weak quantum-classical transition in a one-dimensional, open
system with a bounded classically chaotic Hamiltonian, expanding on
the themes of a shorter paper~\cite{us}.  These arguments are
topological in nature and should be generic for compact,
one-dimensional hyperbolic regions, as well as for unbounded systems
which stretch and fold in a manner analogous to bounded chaotic
systems, unlike other studies which focus on calculations for a
particular system of interest~\cite{others}.  We show how the
classical limit is recovered via two parallel processes.  First,
environmental noise modifies chaotic classical phase space topology by
terminating the production of small scale (late-time) structures.
(This behavior has some parallels with recent numerical studies of a
chaotic advection-diffusion problem with a periodic velocity field, as
will be discussed later~\cite{Voth}).  Second, in the quantum picture,
environmental noise acts as a {\em regulator}, attenuating nonlocal
contributions to the semiclassical wavefunction, and, thereby,
stabilizing a local semiclassical approximation from the pathologies
which a classically chaotic system typically generates, so that it can
now be associated with a noise-modified (smoothed) classical phase
space geometry.  As a consequence of these processes, the local
semiclassical approximation becomes stable at long times, allowing
classical and quantum open systems to be brought into dynamical
agreement at the level of {\em distribution functions}, rather than
the {\em trajectory} level agreement one obtains from conditioning due
to measurements.

The above arguments are very general and apply to a wide class of open
systems.  The key philosophy of our approach is that, for a
classically chaotic system, correspondence is inseparable from some
notion of measurement or environmental coupling. We investigate the associated 
open-system quantum/classical agreement by employing the Wigner
representation of the quantum density matrix and comparing it to the
classical phase space distribution function, an approach with certain
mathematical and formal advantages~\cite{Berry2}. We then utilize this
analysis to elucidate the mechanism by which the agreement occurs, as
well as derive a timescale after which the agreement becomes stable.

Using numerical simulations, we demonstrate the existence of the weak
QCT for the Duffing oscillator and place it in the context of other
numerical studies. Many of the detailed features of the weak QCT for
the Duffing oscillator can be explained and predicted by our
theoretical framework.  We will begin, however, by briefly reviewing
the semiclassical and classical limits of closed nonlinear systems in
the Wigner representation, emphasizing why they disagree with their
associated classical distribution functions at short times and fail as
$t\rightarrow\infty$.  For additional background on this topic see
Ref.~\cite{BenPhD}.

\section{The Nonlinear Classical Limit in Phase Space} 

The Wigner function, $f_{W}(q,p,t)$, is a representation of the
quantum density matrix operator, $\hat{\rho}$, in a $c$-number phase
space~\cite{Wig1}.  Along with the analogous classical phase space
distribution function, $f_{C}(q,p,t)$, we use it to compare the
dynamics of open quantum and classical systems.  Using the Wigner
function as a tool for studying the quantum-classical transition is
conceptually and practically advantageous.  It allows one to compare
classical and quantum dynamics in phase space (though there are
pitfalls one must be aware of), rather than trying to compare, say,
wavefunctions in ${\bf L}^{2}$ to classical trajectories.  More
importantly for our purposes, the theory of semiclassical
approximations can be directly tied to the evolution of classical
curves in phase space, making it easier to visualize the extent to
which quantum and classical dynamical evolutions agree~\cite{Berry2}.
For a classically chaotic system, distribution functions can also give
a clearer sense of global phase space topology, allowing one to
examine the extent to which dynamical agreement over an entire compact
hyperbolic region of interest is achieved.

The Wigner representation of an operator, $\hat{A}$, is defined as:
\begin{equation}
A_{W}(q,p,t)=\int_{-\infty}^{\infty}dX \mbox{ }e^{-ipX/\hbar}\langle
q+\frac{X}{2}|\hat{A}|q-\frac{X}{2}\rangle.    
\label{wigdef} 
\end{equation}
The Wigner function is the Wigner representation of the general mixed
state density operator
$\hat{\rho}=\sum_{i}c_{i}|\psi_{i}\rangle\langle\psi_{i}|$ yielding
\begin{eqnarray}
f_{W}(q,p,t) & = &
\frac{1}{2\pi\hbar}\int_{-\infty}^{\infty}dX
\exp\left(\frac{-ipX}{\hbar}\right)\times   
\nonumber\\ 
&&\sum_{i}c_{i}\psi_{i}(q+\frac{X}{2},t)\psi_{i}^{*}(q-\frac{X}{2},t).
\end{eqnarray}
It follows that, for any operator,
\begin{eqnarray}
\langle\hat{A}\rangle &&=\mbox{Tr}(\hat{A}\hat{\rho})\nonumber\\
&&=\int_{-\infty}^{\infty}dq\int_{-\infty}^{\infty}dp\mbox{ }
A_{W}(q,p,t)f_{W}(q,p,t). 
\end{eqnarray}
Any classical quantity, $q^n p^m$, can be associated with a quantum
operator, via the Weyl ordering
\begin{equation}
\frac{1}{2^n}\sum_{r=0}^{\infty}\hat{q}^{n-r}\hat{p}^r.
\end{equation}
Thus one can compute averages of any classical quantity in the Wigner
picture.

Unlike a classical phase space distribution function, the Wigner
function is only a quasiprobability distribution, as it can take on
negative values.  This condition also implies that the Wigner function
cannot generally be used as a conditional probability distribution and
is bounded by $\pm (\pi\hbar)^{-1}$, which prevents it from being a
delta function in phase space at finite $\hbar$, and, therefore,
prevents it from representing a classical trajectory~\cite{Wig2}.  The
classical phase space distribution is a true positive definite
probability distribution capable of determining averages over
arbitrarily small phase space regions, whereas the degree to which a
Wigner function can capture a local average depends on whether the
region being integrated over is receiving strong quantum interference
effects from locations outside of the integrated region.  

The equation of motion for the Wigner function is given by the Wigner
representation of the equation of motion for the density operator:
\begin{equation}
\frac{\partial f_{W}}{\partial t}=\hat{L}_{C}f_{W}+\hat{L}_{Q}f_{W},
\end{equation} 
where the classical Liouville operator 
\begin{equation}
\hat{L}_{C}\equiv -p\partial_x+\frac{\partial V}{\partial x}\partial_p
\end{equation} 
and the quantum correction
\begin{equation}
\hat{L}_{Q}\equiv\sum_{n\geq1}\frac{\hbar^{2n}(-1)^n}{(2^{2n}(2n+1)!)}
\partial_x^{2n+1}V\partial_p^{2n+1}.
\end{equation}
The form of this evolution equation suggests an intuitive, but
misleading, interpretation of the how the classical limit is
achieved~\cite{LesBook}.  In the equation of motion, $\hbar$ only
appears in the $\hat{L}_{q}$ term.  So it is tempting to suggest that,
as $\hbar \rightarrow 0$, the ``quantum contributions'' to the
evolution of the Wigner function likewise decrease.  However, all of
the momentum derivative terms in both $\hat{L}_{Q}$ and $\hat{L}_{C}$
are proportional to $\hbar^{2n}\partial_p^{2n+1}f_{W}$.  Since, by
definition,
\begin{equation}
f_{W}\sim\exp\left(\frac{ipX}{\hbar}\right),
\end{equation} 
after one takes the appropriate momentum derivatives, it is clear
that, like the wavefunction, $f_{W}$ is $O(\hbar^{-1})$ to leading
order in $\hbar$.  One can never expect quantum corrections to
smoothly disappear as $\hbar$ is decreased due to this essential
singularity, which produces increasingly rapid oscillations as
$\hbar\rightarrow 0$, and will keep the Wigner function from tending
to a positive distribution.  To eliminate the rapid oscillations, one
often introduces an ad hoc filter, as in the case for the Husimi-type
Gaussian filters~\cite{Hus,kaw}.  This can forcibly produce
positive-definite distributions but lacks an underlying
dynamical justification.

The formal study of the classical limit in phase space begins by
constructing a semiclassical Wigner function from an underlying
semiclassical wavefunction.  The semiclassical wavefunction is the
singular $O(\hbar^{-1})$ and constant part of a general wavefunction
in the $\hbar \rightarrow 0$ limit~\cite{Books}.  In this sense, any
small $\hbar$ view of the classical limit must focus on the
semiclassical regime since the semiclassical wavefunction {\it is} the
irreducible part of the wavefunction in this limit.  The standard
presentation tends to view this process as simply representing the two
lowest order terms in a perturbation series for the phase of the
wavefunction.  However, the higher order terms in this series, in
addition to being notoriously difficult to calculate, are rarely
useful.  The remaining terms can be thought of as a vanishing,
$O(\hbar)$ error, and not as a series of higher order terms waiting to
be explicitly calculated~\cite{Maslov}.  

Most importantly, a semiclassical wavefunction is directly associated
with the evolution of classical phase space curves.  The formal
procedure constructs an initial wavefunction from an $N$-dimensional
Lagrangian manifold embedded in a $2N$-dimensional phase
space~\cite{Maslov}.  In this paper, phase space is two-dimensional
and so the associated Lagrangian manifold studied is a curve, which
will be one of a family of phase space curves parametrized by the
continuous parameter $\gamma$, as elucidated in Ref.~\cite{Berry2}.
An initial semiclassical wavefunction associated with the curve
$\gamma$ will have the form:
\begin{equation}
\psi(q,0;\gamma)=A_{0}(q;\gamma)
\exp\left[\frac{i}{\hbar}S_{0}(q;\gamma)\right]  
\label{psi0} \end{equation} where $A_0(q;\gamma)$ and $S_0(q;\gamma)$
are real-valued functions.  For simplicity of presentation, we will
assume that this initial curve has a single momentum value associated
with each position.  Relaxing this assumption would result in a
slightly more awkward presentation, but would not alter its substance.
The above form naturally induces a Lagrangian curve in phase space if
the associated momentum has a well-defined classical limit.  Namely,
\begin{equation} \lim_{\hbar\rightarrow
0}i\hbar\frac{\partial}{\partial
q}\psi(q,0,\gamma)=A_0(q,\gamma)\frac{\partial S_0(q,\gamma)}{\partial
q}\equiv p(q,\gamma).  \end{equation}

The evolving classical curve will typically develop turning points,
which can result in multiple momentum values for a given position.
This will certainly be the case for the highly nonlinear systems
addressed here.  As a consequence of this folding, one assigns a
new action to each branch of the curve as it evolves, as described in
Ref.~\cite{Berry2}.  The action at time $t$ for the $j$-th path is
then given as
\begin{eqnarray}
S_{j}(q,t;\gamma)&=&S_{0}(q_{0j};\gamma)+
\int_{q_{0j}}^{q}dq'p_{j}(q',t;\gamma)\nonumber\\ 
&&-\int_{0}^{t}dt'H(q_{0j},p_{0j}(q_{0j},t';\gamma),t'),
\label{action}
\end{eqnarray}
which yields the semiclassical wavefunction:
\begin{eqnarray}
\psi(q,t;\gamma)&=&\sum_{j=1}^{N}A_j(q,t)
\exp\left(\frac{i}{\hbar}S_j(q,t;\gamma)-\frac{i\pi}{2}\nu_j\right)\nonumber\\ 
&&+O(\hbar),
\end{eqnarray}
where $\nu_{j}$ is the $j$-th Morse index, defined as the number of
times the determinant is equal to zero along the path connecting
$(q_{0j},p_{0j})$ to $(q,p_{j})$.  

By substituting the semiclassical wavefunction into the definition of
the Wigner function, one can construct a geometric interpretation of
the accuracy of a semiclassical analysis.  For the purpose of clarity,
we will assume we are dealing with a pure state density matrix, the
extension to mixed states being straightforward. The semiclassical
Wigner function becomes:
\begin{eqnarray}
&&f_{W}(q,p,t;\gamma)=\frac{1}{2\pi\hbar}\int_{-\infty}^{\infty}dX
\sum_{ij} {\cal A}_{ij}(q,t;\gamma)\times\nonumber\\
&&\exp\bigg[\frac{i}{\hbar}(S_{i}(q+X/2,t;\gamma) 
-S_{j}(q-X/2,t;\gamma)-pX)\nonumber\\
&&-\frac{i\pi}{2}(\nu_{i}-\nu_{j})\bigg],
\end{eqnarray}
where ${\cal A}_{ij}\equiv A_i(q+X/2,t;\gamma)A_{j}(q-X/2,t;\gamma)$.
To get a sense of the primary contributions to this integral as
$\hbar$ is brought to zero and the integrand rapidly oscillates, we
examine the stationary phase condition: 
\begin{equation}
\frac{d}{dX}(S_{i}(q+X/2,t)-S_{j}(q-X/2,t)-pX)=0, 
\end{equation}
where, for clarity, the $\gamma$ parameter is suppressed for the
remainder of the paper.  In the stationary phase approximation, the
Wigner function is separated into a singular stationary part and an
additional $O(\hbar^{1/2})$ oscillatory part~\cite{Bender}.
Therefore, the stationary phases are the most relevant contributions
in the $\hbar\rightarrow 0$ limit, as rapid oscillations become less
significant.  After substituting the expression for the evolved
action, the stationary phase condition becomes: 
\begin{equation}
\frac{1}{2}(p_{i}(q+X/2,t)+p_{j}(q-X/2,t))=p(q,t).  
\end{equation} 
If $i=j$, this is the famous Berry midpoint rule:
$\frac{1}{2}(p_{i}(q+X/2)+p_{j}(q-X/2))=p(q,t)$~\cite{Berry2}.  That
is, the stationary phase contributions at a point $(q,p)$ come from
the average of the momenta on a given solution curve evaluated at the
end of an interval of width $X$ about $q$.
 
If the point $(q,p)$ is particularly close to a curve, $p_{i}(q)$, as
will often be the case when the underlying curve evolves in a chaotic
region of phase space, then the stationary phase points will coalesce,
invalidating the stationary phase method.  Likewise, the WKB
wavefunction itself is not valid near turning points, as the Jacobian
vanishes.  However, these cases are remedied by the uniform
approximation, which yields a symmetric Airy function, rather than
sinusoidal, behavior~\cite{uniform}.  Therefore, if $(q,p)$ is too
close to a given branch, the expression for the semiclassical
contribution for that branch should be replaced by the uniformized
form.  The only remaining problem which can invalidate the expression
is the appearance of catastrophes when $(q,p)$ is a focal point of a
curve, which can be dealt with analytically, as was also studied by
Berry, but is outside the realm of this paper.  Close to the classical
curve $p_{i}(q)$, the uniformized WKB approximation for the Wigner
function has an Airy ``head'' of width $\sim\hbar^{2/3}$ and peak
height $\sim\hbar^{-2/3}$. In this limiting case, the uniform
approximation can be further simplified and written in the form of a
``transitional approximation'' which is valid only very near
$p_{i}(q)$. Remarkably, the $\hbar\rightarrow 0$ limit of the
transitional approximation is indeed a classical delta function, which
allows the Wigner function picture to give a physically clearer
presentation of the classical limit.

The semiclassical quantum Wigner function goes through three phases in
its evolution, if the underlying classical dynamics is chaotic, as
laid out by Heller and Tomsovic~\cite{heller}.  During the (very
short) first phase, if the initial condition used is that of a
classical distribution, there will be little disagreement between the
quantum and classical evolutions. This is followed by a second phase,
where the semiclassical approximation reproduces the wave function
dynamics, but is distinctly nonclassical.  At a longer timescale,
proportional to inverse powers of $\hbar$, the semiclassical
approximation fails, as the distance between classical manifolds
becomes so close that the cumulative interference cannot be locally
ascribed to any given curve.  So, in the first, classical regime,
there is little interference.  In the second, semiclassical regime
there is some, possibly strong, quantum interference, but it is in the
form of local fringing about classical curves.  In the final, fully
quantum phase, there is strong global interference, and local
classical manifold evolution is of little relevance to the quantum
propagation.

There are two sources of quantum interference in phase space: local
Airy ``shadows'' of the short wave classical curve and nonlocal
contributions from multiple curves, the latter being more problematic
if we wish a weak QCT to hold.  In order to maintain a stable
classical limit for a classically chaotic system, it must be possible
to keep a system in a more or less classical regime (analogous to, but
not the same as the first regime discussed above), allowing only a
small admixture of local interference effects. We show below how such
a stable classical limit arises in open systems, via the same physical
process that simultaneously leads to a smoothing of the classical
phase space geometry.

\section{Open Systems and Measurement}
To model the interaction between a subsystem and its measuring device
we choose the form of an unconditioned continuous position
measurement.  This provides the minimum level of interaction necessary
to bring quantum and classically chaotic dynamical systems into
(approximate) agreement with one another at the level of distribution
functions.  The model of a conditioned continuous position measurement
(i.e., evolution of the system density matrix taking the results of
measurement into account) is given by the following master
equation~\cite{kurtdan}:
\begin{eqnarray}
d\rho&=&-{\frac{i}{\hbar}[H,\rho]+k[X[X,\rho]]}dt\nonumber\\
&&-\frac{\sqrt{\bar{k}}}{2}([X,\rho]_{+}-2\rho\langle X\rangle)dW,
\label{rhosme}
\end{eqnarray}
where the observed measurement record is given by
\begin{equation}
dy=\langle X \rangle dt+\frac{1}{\bar{k}}dW.
\end{equation}
In the above equation $\langle X \rangle=Tr(\rho X)$, $dW$ is the
Wiener measure [$(dW)^{2}=dt$], $k$ represents the strength of the
interaction between the subsystem and the measuring apparatus and
$\bar{k}$ measures the rate at which information about the system is
being extracted. The fractional measure of extracted information is
given by the efficiency of the measurement
$\eta\equiv{\bar{k}}/8k$. The first term in Eqn.~(\ref{rhosme}) is
just the unitary evolution for the closed system, the second is a
diffusive term arising from quantum backaction, and the third
represents the conditioning due to the measurement.
 
The conditioned evolution can localize the state about the measured
position value; the extent of this localization (proportional to
${\bar{k}}$) must however be tempered by the associated increase of
backaction noise (concomitant increase in $k$). Nevertheless,
inequalities can be derived that show under what conditions both of
these conflicting effects can be reconciled and agreement between
classical and quantum dynamics achieved at the level of
trajectories~\cite{Tan,SSJ} -- the strong form of the QCT.  

If one averages over all obtained measurement records, one obtains the
master equation for an unconditioned evolution:
\begin{equation}
d\rho=-{\frac{i}{\hbar}[H,\rho]+k[X[X,\rho]]}dt.
\label{uncon}
\end{equation}
This evolution can also be achieved by setting the efficiency of the
measurement, and, therefore, $\bar{k}=0$.  Once one does so, the
localization inequalities which characterize the strong form of the
QCT fail, showing the inability of the weak QCT to capture trajectory
level chaos and the need for the distribution function approach
employed here.  The evolution equation is the same as that for the
Caldeira-Leggett model in the weak coupling, high temperature
approximation~\cite{CL}. The key point here is that while the
conditioning term is absent, the backaction term remains. This is very
different from the classical case, where averaging over measurements
simply gives back the closed-system Liouville equation, thus
highlighting the contrast between the active nature of quantum
measurements versus the passive nature of classical measurements.

The master equation (\ref{uncon}) is the starting point in our
analysis of the weak QCT utilizing the Wigner function. In the Wigner
representation, this equation becomes 
\begin{equation}
\frac{\partial f_{W}}{\partial
t}=\hat{L}_{C}f_{W}+\hat{L}_{Q}f_{W}+D\frac{\partial^{2}f_{W}}{\partial
p^{2}}, 
\label{unconwdf}
\end{equation}
where the diffusion coefficient $D=\hbar^{2}k$.  If we set
$\hat{L}_{Q}=0$, we obtain a dual classical evolution equation, for
the classical distribution function $f_{C}(q,p,t)$: 
\begin{equation}
\frac{\partial f_{C}}{\partial
t}=\hat{L}_{C}f_{C}+D\frac{\partial^{2}f_{C}}{\partial p^{2}}, 
\label{dual eq}
\end{equation}
which, given its form, we will call the dual Fokker-Planck
equation. Note that this Fokker-Planck equation does {\em not}
represent the dynamics of an associated classical observed system.
Here it has two key roles: it represents the classical template for a
semiclassical open-system analysis and also the proper (approximate)
classical limiting form if the weak QCT were to hold. This particular
Fokker-Planck equation is better viewed as simply a classical dual of
the quantum master equation (\ref{unconwdf}), without an independent
physical existence.

A final note on timescale separations is necessary to clarify the
physical situations under which Eqns.~(\ref{uncon}) and
(\ref{unconwdf}) are considered to hold. We are not interested in
imposing initial conditions on the quantum dynamics that have
classical analogs (e.g., Gaussian wavepackets), and then looking for
the emergence of short-time quantum effects. In fact, we acknowledge
the existence of quantum initial conditions explicitly (as in the
numerical simulations of Section VI), and investigate
quantum-classical convergence in the sense of the convergence of
distribution functions as obtained from the quantum master equation
and its classical dual.

At the the same time, we are particularly interested in the dynamics
set by the closed-system Hamiltonian, with minimal influence from the
external environment or continuous measuring process, i.e., the weak
coupling limit. In this limit, we can ignore the dissipative effects
of external couplings (damping due to environment modes and/or
measurement backaction), but consider only diffusive effects, which
remain finite in the weak coupling limit (as in the weak-coupling,
high-temperature Caldeira-Leggett model). There are two timescales
associated with these statements. The first, $t_{relax}$, is the time
taken for the system to relax to a thermal state, (or nonequilibrium
steady state depending on the circumstances) and is typically
controlled by the matching of energy exchange as set by the
dissipation and diffusive channels. The second, $t_{diff}$, is the
diffusive heating timescale which, in the absence of dissipation,
leads to continuous heating of the system. Since at late times, when
dissipative effects would be expected to occur, this heating is
unphysical, it is clear that our analysis assumes $t\ll
t_{diff}$. Therefore, we are interested in the dynamics of open
quantum systems on intermediate timescales, longer than the system
dynamical timescales, yet far from the (asymptotic) timescales relevant
for close to steady-state behavior. All remarks below on ``long-time''
behavior apply to this intermediate timescale and not to some eventual
steady state.

\section{Modification of Phase Space Geometry for a Chaotic
 Subsystem}  

The first step in our analysis is the study of the dual Fokker-Planck
equation. As earlier mentioned, following a semiclassical line of
reasoning, the motivation for this is that the
measurement/environmental interaction modifies the geometry of a
chaotic classical phase space in a manner which can allow dynamical
agreement between classical and quantum systems. The key point is
that, due to the diffusion term, one necessarily sees a termination in
the level at which one can discern the long-time development of fine
structure. The (exponential) long-time development of structure is a
hallmark of classically chaotic systems in a compact space, and, as
discussed in the previous section, leads to disagreement between
classical and semiclassical results, followed by a complete failure of
semiclassical analysis.  But, as these structures are averaged over,
the resulting smoother phase space geometry can be consistent with the
existence of a local semiclassical description.

We will show below that the diffusion term in the Fokker-Planck
equation terminates the development of small scale structures at a
finite time, denoted by $t^{*}$.  At this time, there will be an
associated area, $l_{cl}(t^{*})^{2}$, below which no smaller phase space
structures can be discerned.  To understand the termination of
structure, we consider the Langevin equations underlying the dual
Fokker-Planck equation. These are given by
\begin{equation}  
dq=p dt/m
\end{equation}
and 
\begin{equation}
dp=f(q)dt+\sqrt{2D}dW,
\end{equation}
where $f(q)=-\partial V(q)/\partial q$, $dW$ is the Wiener measure
[$(dW)^{2}=dt$], and $D$ is the noise strength.  Since $D$ is constant, 
one can consistently write $dW=\xi(t) dt$,
where $\xi(t)$ is a rapidly fluctuating force satisfying
$\langle\xi(t)\rangle=0$ and
$\langle\xi(t)\xi(t')\rangle=\delta(t-t')$ over noise averages.

A hyperbolic region of the phase space of a bounded chaotic
Hamiltonian system is foliated by its unstable manifold, which emerges
from the stretching and folding behavior induced when nonperiodic
solution curves are confined to a bounded region.  A trajectory in the
neighborhood of a hyperbolic fixed point will create large scale
structures, due to its exponential growth away from the hyperbolic
point.  As it evolves, since it can only explore the energetically
allowed region, it will fold onto itself and create smaller scale
structures.  For a bounded chaotic region, the curve will eventually
fill the allowed space.  The important consequence for this analysis
is that this filling is done preferentially.  Large scale structures
are initially generated by rapid stretching and are associated with
short timescales.  The smaller scale fine structures are then filled
in afterwards as the system continues to fold on itself and are,
therefore, a late-time feature.

In order to investigate how environmental noise modifies this picture,
we perform a perturbative expansion of the solution curve in the small
noise limit in the neighborhood of a hyperbolic fixed point
$(q_{eq},0)$, where $f(q_{eq})=0$, where $\sqrt{2D}$ is treated as the
small noise parameter~\cite{Gard,Kamp}.  As emphasized in the previous
section, this assumption is physically justified by the argument that
the affected noise scale in phase space should be smaller than that of
the system dynamics.  As discussed in the introduction, the system is
taken to be weakly interacting with its environment to ensure that the
system dynamics are affected only perturbatively. To leading order in
$\sqrt{2D}$, we can therefore separate the dominant systematic
components from the noisy components via $q(t)\approx q_{C}(t)+
q_{N}(t)$ and $p(t)\approx p_{C}(t)+ p_{N}(t)$, leading to the usual
Hamilton's equations for $q_{C}$ and $p_{C}$, and to the coupled
equations $dq_{N}=p_{N} dt/m$ and $dp_{N}=m\lambda^{2}q_{N}dt+dW$,
where $m\lambda^2=\partial f(q_{eq})/\partial q$ defines the local
Lyapunov exponent, $\lambda$.  These have the solution,
\begin{eqnarray}    
q(t)& = & q_{eq}+C_{+}e^{\lambda t}+C_{-}e^{-\lambda t} \nonumber\\ 
&& +\frac{\sqrt{2D}}{2m\lambda}\int_{0}^{t}du \xi(u)
\left(e^{\lambda(t-u)}-e^{-\lambda(t-u)}\right), 
\label{lansol}
\end{eqnarray}
with an analogous expression for $p(t)$.  

To understand the effect of noise on the foliation of the unstable
manifold, one needs to transform from the position and momentum basis
into the stable and unstable directions.  The dimensional scalings
$q'=\sqrt{\lambda m}q$ and $p'=p/\sqrt{\lambda m}$ are introduced so
that the rescaled position and momentum have the same dimensions and
also so that the stable and unstable directions are orthogonal.  An
arbitrary time rescaling, which would give the correct units, would
not guarantee orthogonality.  If we project the solutions for
$q'=\sqrt{\lambda m}q$ and $p'=p/\sqrt{\lambda m}$ along the
stable~(-) and unstable~(+) directions, we find the following
expression for the components of the noisy trajectories evolution in
these two directions:
\begin{eqnarray}      
u_{\pm}(t)&=&{1\over\sqrt{2}}(q' \pm p')\nonumber\\
&=&\sqrt{2\lambda m} C_{\pm}e^{\pm\lambda t}\pm\sqrt{\frac{D}{\lambda m}}
\int_{0}^{t}du\xi(u)e^{\pm\lambda(t-u)}.\\
\label{proj}
\end{eqnarray}

One can now analyze the effects of these noisy trajectories on the
evolution of the distribution function which they unravel.  The
average over all noisy realizations of the displacement in the stable
and unstable directions is given by $\langle
u_{\pm}\rangle=\sqrt{2\lambda m}C_{\pm}e^{\pm\lambda t}$, as expected
from a perturbation in the neighborhood of a hyperbolic fixed point.
More information is found in the second order cumulants.  Whereas, the
stable and unstable directions have variances of $\pm
(D/(2m{\lambda}^{2}))(e^{\pm 2\lambda t}-1)$, the off-diagonal
cumulant is $\langle u_{+}u_{-}\rangle-\langle u_{+}\rangle\langle
u_{-}\rangle=-Dt/(m\lambda)$, displaying the linear spreading
associated with a Wiener process.  In forward time, where the
evolution of a trajectory is determined by the unfolding of the
unstable manifold, this spreading indicates that, as the trajectory
evolves, it will simultaneously smooth over a transverse width in
phase space of size
\begin{equation}
l_{cl}(t)\approx\sqrt{{Dt}/(m{\lambda})}.
\label{width}
\end{equation}  
One is left with a picture of a curve following a classical path in
the unstable direction while carrying small amounts of transverse
noise. In a bounded, compact phase space region, this implies a
termination in one's ability to measure the position and momentum of
the trajectory on a scale smaller than the aforementioned width.  In
other words, the fine structures associated with a chaotic region will
be smoothed over in the averaging process, causing the development of
large scale structures which occur prior to this termination time to
become pronounced.

Given a set of parameters associated with this compact phase space
region, one can estimate the value and scaling associated with the
termination time, $t^*$.  Consider an initially small compact region
of phase space area $u_{0}^2$, then its current phase space ``length''
is approximately $u_{0}e^{\bar{\lambda} t}$, where $\bar{\lambda}$ is
the time-averaged positive Lyapunov exponent. If the trajectory is
bounded within a phase space area $A$, the typical distance between
neighboring folds of the trajectory is estimated by
\begin{equation}
\delta(t)\approx A/(u_{0}e^{\bar{\lambda}t}).
\label{delta}
\end{equation}
This formula only applies once the curve has begun to fold on
itself; given a particular choice of $u_{0}$, one must be careful that
enough time for folding to occur has passed before using the above
equation. One can, in this spirit, estimate a ``folding time'' and
compare it with the eventual computed value of $t^*$ to again insure
that this analysis is self-consistent. In any case, one cannot make
$u_{0}^2$ arbitrarily small when exploring the QCT because the
uncertainty principle sets a lower bound on phase space area. Note
that the length of a long timescale -- long compared to the dynamical
timescale --is also implied by the appearance of the time-averaged
Lyapunov exponent, $\bar{\lambda}$.

Phase space structures can only be known to within the width specified
by the noisy dynamics, hence there will come a time at which the
rapidly falling scale $\delta(t)$ set by the folding will be smaller
than the slowly increasing filter scale, $l_{cl}(t)$, at which lengths
are averaged over as given by Eqn.~(\ref{width}). The time at which
any new structures will be smoothed over is given by equating
Eqns.~(\ref{width}) and (\ref{delta}), to yield
\begin{equation}
\sqrt{\frac{Dt^{*}}{(m \bar{\lambda})}} = 
\frac{A}{u_{0}}\exp(-\bar{\lambda}t^*),
\label{nscale}
\end{equation}
a simple transcendental equation for $t^*$. Typically, $t^*$ is
expected to be significantly larger than $1/\bar{\lambda}$, as there will 
usually be many foldings before the filtering becomes effective.  In this 
case a simple iterative procedure can be used to find the approximate
solution,
\begin{equation}
t^*\simeq \frac{x_0}{2\bar{\lambda}}\left[1-\frac{\ln(x_0)}{1+x_0}\right],
\label{tstar}
\end{equation}
where $x_0=\ln(2m{\bar{\lambda}}^2A^2/(Du_0^2))$.

After the time $t^*$ no new structures will be discerned, since they
will be smaller than the averaging scale set by the noisy dynamics.
This implies the existence of a phase space area $l_{cl}(t^{*})^2$
below which phase space structures are smoothed over.  As a result, 
the dual Fokker-Planck equation for a chaotic system is such that we can only
discern large scale structures (small and large being relative to the
cutoff $l_{cl}(t)$ produced prior to $t^*$). When constructing a
classical limit for the open quantum evolution (\ref{unconwdf}), we
now only have to capture the larger, short-time dynamical features and
not the full chaotic evolution of the classical Liouville equation
with its -- from a quantum perspective -- small-scale pathologies.

Before proceeding further, we mention an analogous situation in
studies of chaotic advection-diffusion in fluid dynamics.  The
evolution equation for the concentration density, $c({\bf x},t)$, of a
set of particles diffusing in a fluid without sinks or sources is
given by
\begin{equation}
\frac{\partial c}{\partial t}+\nabla c \cdot {\bf v}=\kappa\nabla^2 c,
\end{equation}
where ${\bf v}({\bf x},t)$ is the velocity field of the tracer
particles.  This matches the classical Fokker-Planck equation studied
here, if one sets
\begin{equation}
{\bf v}=\left(\frac{p}{m},-\frac{\partial V}{\partial q}\right)
\end{equation}
and the gradient is taken with respect to $q$ and $p$.  Diffusion, in
our case, is only with respect to $p$.  The phase space distribution
function is then regarded as the concentration of particles in phase
space in a given region, which is certainly an appropriate
interpretation.  A numerical analysis performed in Ref.~\cite{Voth}
showed results similar to our predictions where, for a certain vale of
$\kappa$, equivalent to $D$ in our case, the evolution converged to a
stationary pattern at a finite time, with only residual diffusion
afterwards.  The final pattern was termed an inertial manifold and
related to the unstable manifold.  Following this, the existence of
such a manifold beyond a critical $\kappa$ value was demonstrated
analytically~\cite{LH}.

This suggests that the qualitative classical analysis provided here
might be made more rigorous. The analysis of Ref.~\cite{LH} relied
significantly on applying periodic boundary conditions to the
concentration evolution and exploiting the resulting gaps in the
spectrum of the Laplacian. The open boundary conditions relevant to
this paper, however, appear to preclude such an
approach. Nevertheless, qualitative similarities do exist and the two
fields may well inform each other in future.  (Of course, this is a
purely classical analysis and does not bear directly on the quantum
evolution, except implicitly since semiclassical evolution tracks the
classical manifold structure.)
 
\section{Semiclassical Analysis for an Open Chaotic System}

We now turn to the semiclassical analysis of the open system master
equation (\ref{unconwdf}) in order to estimate the conditions under
which a weak QCT might exist. We begin by rewriting the semiclassical
Wigner function in the weak noise limit utilized in the previous
section.  In this limit, the classical action is modified to
$S(q,t)\approx S(q_{C},t)-\sqrt{2D}\int_{0}^{t} dt \xi(t)q_{C}(t)$, as
in Ref.~\cite{Kos}.  The first term will evolve classically, as
discussed in Section III, as will the position coordinate which
appears in the second term.  If we insert the above semiclassical
action into the expression for the Wigner function we get the
following result:~
\begin{eqnarray}
f_W(q,p,t) &=& \int dX\mbox{ } \frac{e^{-ipX/\hbar}}{2\pi\hbar}
\sum_{i,j}^{N}{\cal A}_{ij}\nonumber\\  
&&\exp\left(-\frac{i}{\hbar}X\sqrt{2D}\int_0^t dt'\mbox{
  }\xi(t')\right)\times\nonumber\\ 
&&\exp\bigg(\frac{i}{\hbar}(S_{Ci}(q_{C+},t,P)-S_{Cj}(q_{C-},t,P))\nonumber\\
&&-\frac{i\pi}{2}(\nu_i-\nu_j)\bigg),
\end{eqnarray}
noting that, since the amplitude is a second derivative and the noisy 
perturbation is linear, noise only effects the action to lowest order.  

If we next average over all noisy realizations, the following
suggestive expression for the noise averaged semiclassical Wigner
function is obtained:
\begin{eqnarray}
&& \int dX\mbox{ } \frac{e^{-ipX/\hbar}}{2\pi\hbar}
\exp\left(-\frac{Dt}{\hbar^2}X^2\right) \sum_{i,j}^{N}{\cal
  A}_{ij}\times\nonumber\\ 
&&\exp\bigg(\frac{i}{\hbar}(S_{Ci}(q_{C+},t,P)-S_{Cj}(q_{C-},t,P))\nonumber\\
&&-\frac{i\pi}{2}(\nu_i-\nu_j)\bigg).
\end{eqnarray}
The only alteration to the expression for the semiclassical
wavefuntion to lowest order in the noise strength is the appearance of
a new Gaussian term.  The presence of noise acts as a dynamical
low-pass Gaussian filter of semiclassical phases, attenuating large
$X$ contributions.  For any solutions to the above equation, phases
will be suppressed which have wavelengths greater than
\begin{equation}
X\approx\hbar/\sqrt{Dt}.
\label{filter}
\end{equation}  
These are the long, nonlocal ``De Broglie'' wavelength contributions
to the semiclassical integral, the very sort of contributions
previously identified as being particularly problematic in terms of
obtaining a weak QCT. The filter prevents the integral from becoming
overwhelmed by long range contributions as stretching and folding
occurs which can lead to disagreement with classical results, as well
as the eventual failure of the approximation.

We now combine the above with the classical result from the last
section. It is seen that the diffusion causes two effects: suppression
of nonlocal phases in the semiclassical integral beyond a certain
scale given by Eqn.~(\ref{filter}) and a smoothing of the dual
classical phase space over fine structures smaller then a scale given
by Eqn.~(\ref{nscale}).  Each of these effects overcomes the two
semiclassically identified difficulties associated with a weak QCT for
chaotic systems: the Wigner function is no longer dominated by
nonlocal contributions and also does not need to track, nor does it
receive interference from, very fine scale structures.  From these two
scales we should, therefore, be able to set a (semiclassical) criteria
for the existence of a weak QCT for a bounded one-dimensional chaotic
system.  Physically, the local semiclassical approximation is valid
when the primary contributions to the semiclassical integral at a
given point $(q,p)$ come from the local branch of the trajectory on
which the point is located.  This will occur only when the scale at
which local classical smoothing occurs matches or exceeds the
filtering scale for semiclassical phases.  When this occurs the
nearest possible branch which is capable of delivering nonlocal
interference effects will have those effects filtered within the
semiclassical integral.  As a result one can recover the usual short
wave semiclassical picture of a trajectory ``decorated'' only by local
interference fringes.

More specifically, if we rescale the filtering
condition~(\ref{filter}) in phase space units
\begin{equation}
l_q\approx\frac{\sqrt{m\bar{\lambda}\hbar}}{\sqrt{Dt}},
\label{lq}
\end{equation}
then the semiclassical criterion for the weak QCT is given by
$l_q(t)<l_{cl}(t)$. The quantum scale $l_q$ decreases with time, as the
noise filtering, beginning with ``fast'' phase space oscillations
(interference due to far-separated features in phase space), reaches
down to ever ``slower'' interference scales (due to small-scale phase
space features). If small-scale classical structures continued to be
generated in phase space at a rate outstripping the decrease of $l_q$
with time, the QCT would not occur. Due to the presence of noise,
however, classical small-scale structure does not grow exponentially,
but is eventually cut off by $l_{cl}$, which grows with
time. Therefore, $l_q$ and $l_{cl}$ must cross each other, and this
point defines the weak QCT timescale, $t_{qc}$. For $t>t_{qc}$, the
classical structures are large enough that the noise filtering is
effective in smoothing over the associated interference terms.

The weak QCT timescale follows from equating Eqns.~(\ref{width}) and
(\ref{lq}) for $l_{cl}$ and $l_q$, respectively,
\begin{equation}
t_{qc}\approx m\hbar\bar{\lambda}/D.
\label{tqc}
\end{equation}
Using Eqn.~(\ref{width}) once again, one finds that this condition is
nothing but $l_{cl}^2(t_{qc})=l_q^2(t_{qc})\approx \hbar$, which would
have been suggested by basic intuition. Following from the discussion
above, $t_{qc}$ can also be interpreted as the timescale beyond which
a semiclassical approximation becomes stable for an open quantum
system.  After this time, classical dynamics should approximate
quantum dynamics sufficiently.

Note that the two timescales discussed so far, $t^*$ and $t_{qc}$,
scale very differently with the diffusion coefficient, $D$. Whereas,
$t^*\sim \ln(1/D)$, $t_{qc}\sim 1/D$, implying that the timescales are
far-separated in the small $D$ (weak noise) limit, where typically,
$t_{qc}\gg t^*$. It is possible, however, to have $t_{qc}<t^*$ even at
modest values of $D$. The physical interpretation of these two
possible situations is as follows. As discussed previously, the
timescale $t^*$ sets the ``freeze-out'' of classical phase space
structures, but it is possible to have a weak QCT occur on either side
of the freeze-out. When $t_{qc}>t^*$, even though the large-scale
classical phase space template is relatively fixed, small-scale
discrepancies will exist between the quantum and classical
distributions at least until $t\sim t_{qc}$. Though, in this case, the 
classical filtering will have terminated the development of classical 
structures at $t^*$, some time must still elapse before interference between 
branches has been sufficiently filtered.  On the other hand, when
$t_{qc}<t^*$, the weak QCT can occur while the large-scale classical
phase space structures are still evolving since the classical freeze-out 
has not yet taken place.

\section{Numerical Simulations}
The analysis in the preceding sections has helped to establish a set
of criteria which, once met, allow the existence of a weak QCT for
classically chaotic systems. Given their somewhat heuristic nature, it
is important to examine these predictions numerically. In the quantum
evolution, once the inequalities are satisfied, noise will filter
nonlocal quantum interference between the surviving large scale phase
space structures, so large scale coherences should not be present.  If
not, one will essentially see a global phase space diffraction
pattern, with large-scale coherences persisting between all parts of
the bounded phase space region, an example of which is shown in Figure 1.
\begin{figure}[htbp]
\begin{center}
\includegraphics[width=3.6in]{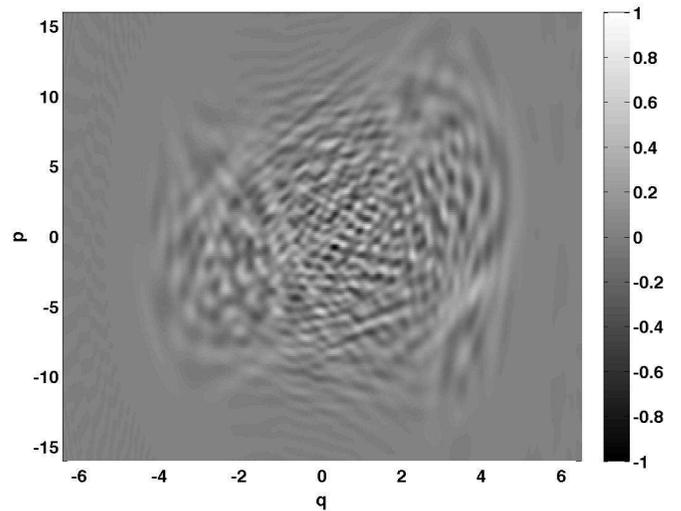}
\caption{\label{fig:Pud} Phase space rendering of the Wigner function
  for the Duffing system at time $t=314~\mbox{periods of driving}$.
  The nonlocal interference is significant and cannot be associated
  with specific classical structures.  This plot is taken at a
  relatively small $D$ value ($10^{-4}$) for resolution purposes.  The
  value of $\hbar$ is set equal to 1 in order to clearly demonstrate
  this effect.}
\end{center}
\end{figure}

The most direct numerical test is a close examination of the time
evolution of both the classical and quantum distribution functions for
the quantum and dual classical evolutions.  In this manner one can
examine whether, over the expected timescales as predicted in the
previous section, the expected phase space features are present for
the template classical distributions and quantum Wigner functions.

A direct examination is necessary as other, seemingly logical
measures, can sometimes be misleading. For instance, looking at
expectation values is not always helpful.  Typically, one would sample
a set of lower order moments, and follow their expectation values for
a desired amount of time.  However, any classically chaotic system
will have, over time, an infinite set of non-vanishing moments, as
will any non-Gaussian quantum system~\cite{gaussian}.  That said, one
might attempt to argue that the effect of these higher-order moments
may well be negligible.  Even if this were the case, however,
numerical simulations of chaotic systems have failed to find
well-defined break times at which even lower-order moments permanently
separate, and expectation values can agree well for surprisingly long
timescales and even without the presence of environmental
noise~\cite{heller}. Other measures, such as suppression of the
integrated negativity of the Wigner function, also are not necessarily
signatures of quantum-classical correspondence, as shown in
Ref.~\cite{hjmrss}.  Once the amount of negativity is eliminated via
environmental decoherence, it was shown by these authors that the
quantum evolution of a system may still disagree with its classical
counterpart.

\subsection{Numerical Methods}

Numerical solutions of the quantum master equation (\ref{unconwdf})
for the Wigner function and of the corresponding dual classical
Fokker-Planck equation were carried out using a split operator
spectral method implemented on parallel supercomputers~\cite{FF}. The
spectral method is particularly well-suited to high spatial resolution
simulations where spatial structure is cut off above some given
wavenumber -- this is the case here for both the quantum and dual
classical evolutions.

The time-stepping strategy is the same as that in analogous classical
symplectic integrators. Suppose the time evolution of a function,
$f(t)$, satisfies the operator equation:
\begin{equation}
\frac{\partial f}{\partial t}=(\hat{L}_{A}+\hat{L}_{B})f,
\end{equation} 
where the separate evolutions given by $\hat{L}_{A}$ and $\hat{L}_{B}$
can be implemented exactly. The exact solution to this equation is
given by: 
\begin{equation}
f(t)=e^{(\hat{L}_{A}+\hat{L}_{B})t}f(0).
\end{equation}
Since $\hat{L}_{A}$ and $\hat{L}_{B}$ do not commute in general, the
fact that the individual evolutions are known exactly is not of
direct use. An integration scheme for a small timestep $\Delta t$ can
be constructed simply, however, using the Campbell-Baker-Hausdorff
theorem: 
\begin{equation}
f(\Delta t)\approx e^{(\frac{\Delta t}{2}\hat{L}_ A)}e^{(\Delta t
\hat{L}_B)}e^{(\frac{\Delta t}{2}\hat{L}_A)}f(0)+O(\Delta t^3). 
\end{equation}
With the assumption that the exponentiated operators can be applied
exactly, this method is accurate to second order in $\Delta t$.  The
third order correction term is 
\begin{equation}
\frac{1}{24}(\Delta t)^3 [\hat{L}_A +2\hat{L}_B,[\hat{L}_A,\hat{L}_B]]f(0),
\end{equation}
which can be evaluated to estimate the accuracy of the approximation. 

In the present case, the evolution operator is
$\hat{L}_{cl}+\hat{L}_{q}+D\partial^{2}_{p}$ for the Wigner evolution
and is the same, but with $\hat{L}_q=0$ for the dual classical
evolution. We split this into three operators, the ``stream'' operator
$-(p/m)\partial_{q}$, the ``kick'' operator proportional to potential
derivatives, which differs for the classical and quantum cases, and
the momentum diffusion operator.  As each piece involves either
derivatives of position or momentum, but not both, the individual
operators can be easily evaluated using a fast Fourier transform.  The
split-operator method preserves the unitarity of evolutions when $D=0$
and given a sufficient number of grid points in the spatial and
momentum directions -- satisfying associated Nyquist conditions -- the
operators can be evaluated at each timestep with essentially no
spatial discretization error.

The typical mesh used over phase space consisted of 4096 by 4096 grid 
points.  This size was determined by our need to resolve the bounded
{\it classical} phase space portrait for the amount of time necessary
to show long range agreement between the classical and quantum
evolutions.  If $D=0$, the classical phase space will be chaotic, and
the system can only be explored for short times, before which
structures begin to proliferate on scales smaller than the area
defined by the grid spacing.  The addition of an environmental
interaction, as demonstrated in the theoretical section, prevents
structures from forming on infinitely small scales.  This makes it
possible for the classical evolution to converge as resolution
improves.  The aforementioned grid-size is the one for which
convergence was achieved for the systems we studied, and was derived
empirically. Convergence for the quantum evolution is determined by
the smallest scales -- $\delta x=\hbar/P$ in space and $\delta
p=\hbar/L$ in momentum -- present in the Wigner function ($L$ and $P$
are the scales of the system boundaries in length and momentum,
respectively). Thus, a typical mesh spacing can be fixed without
regard to the strength of the environmental interaction. For our
investigations, this required less resolution than in the classical
phase space and, therefore, the dual classical evolution dictated the
grid-size for the numerical simulations.

\subsection{Duffing Oscillator}

The particular potential chosen for study was the chaotic Duffing
oscillator with unit mass: $H(q,p,t)=p^2/2+Bx^4-Ax^2+\Lambda
x\cos(\omega t)$.  The evolution was evaluated for the set of
parameters $A=\Lambda=10$, $B=0.5$ and $\omega=6.07$.  In this
parameter regime, the system is strongly chaotic, with an average
Lyapunov exponent of $\bar{\lambda}=0.57$ that is relatively uniform
over the hyperbolic phase space region~\cite{Ball}.  The size of the
bounded phase space region, which is $A$ in our calculations, is
approximately $270$ units of action. The hyperbolic region of the
system's bounded motion is generated by the homoclinic tangle of a
single hyperbolic fixed point and the stable regions are relatively
small.  Consequently, the unstable manifold associated with the
hyperbolic point completely characterizes the chaotic region and
provides an ideal test for the theory developed in this paper for
bounded hyperbolic regions.

These parameters were chosen, not only because they provide
appropriate testing conditions for theory, but also because their
classical dynamics have been well studied in Ref.~\cite{Ball} and
elsewhere.  As a result, one can estimate the values of the quantities
of interest, e.g., $t_{qc}$, with the system parameters, such as
$\bar{\lambda}$, fixed at some canonical values.  In addition, one
also must be careful to choose a value of $\hbar$ which is not so
large that the initial conditions are well outside the bounded region.
Of course, choosing a value of $\hbar$ of the same order of magnitude
as the bounded region or greater, would also invalidate the argument.
One also does not want to choose $D$ values which are very large
compared to those at which the transition is predicted to occur, as
extreme $D$ values, while inducing quantum-classical correspondence,
may wash out any intrinsic system dynamics.

We will principally focus on the case where $\hbar=0.1$ for a variety
of practical reasons.  The value of $\hbar=0.1$ turns out to be
convenient for these purposes: the critical $D$ value predicted is
small, but not too small that it is below computational resolution,
and it also allows a wide range of $D$ values to be studied without
smearing out the system dynamics.  This value of $\hbar$ was used in
Ref.~\cite{SalKo} which motivated much of this research, and which
confirms that a weak transition will occur for this value.  Still, the
additional set of $\hbar$ values, $\{0.01, 0.5, 1, \sqrt{2}, 3, 5, 10,
20\}$, were studied, and all revealed similar results, though some,
such as $\hbar=0.01$ and $\hbar=5$, had compromised dynamical ranges,
while $\hbar=10$ and $20$, were too large to be of practical interest.
The results presented in depth in this section for $\hbar=0.1$ should,
therefore, be thought of as emblematic of all cases studied.  For a
given trial, $\hbar$ and $D$ were held fixed.

We use the same normalized initial conditions for both dual
classical and quantum evolutions, as we are trying to see the degree
to which the two evolutions follow each other.  In the numerical
simulations, the typical condition was a superposition of two
Gaussians, since a classically unacceptable initial condition would
better illustrate the suppression of interference effects.  Other
conditions were also tried and compared, with analogous results.

The choice of $D=0.001$, yields the estimates $t^*=15.02$ and
$t_{qc}=57$.  For this case, the development of large scale classical 
structures should terminate before quantum and classical agreement occurs.  
A larger value for the diffusion coefficient, $D=0.01$, gives $t^*=13.1$ 
and $t_{qc}=5.7$. Here the transition occurs just before the termination of 
classical structure.  Because of the quantum nature of the initial condition, 
in the estimation of $t^*$, we have set $u_0^2=\hbar$.

We now set out to test these predictions via numerical simulations. We
first compare the classical and quantum evolutions at late times in
order to establish whether or not a quantum-classical transition in
fact occurs as predicted. Comparison of expectation values was helpful
to establish whether the transition had occurred, but final approval
was given only after examining the distribution functions and Wigner
functions directly. Such a comparison is presented in Figure~2.  We
compare cross-sectional slices of the classical distribution function
and quantum Wigner function after $149$ drive periods of the Duffing
oscillator.  These slices are taken along the $p=0$ line.  For
$D=10^{-5}$ very little agreement occurs between the classical and
quantum slices (as expected, since in this case $t_{qc}\sim 5700$).
In fact, the quantum slice still has many negative regions. For
$D=10^{-3}$, in agreement with our order of magnitude estimate for
when a transition should occur, progress has clearly been made.  The
two functions are in average agreement with one another, and, although
there is less agreement on details, there is agreement between the two
on some of the larger phase space feature. At a larger value,
$D=10^{-2}$, this agreement is much improved and overall, the
distributions are much smoother. 

Note that the weak QCT occurs in time when
$l_q=l_{cl}\sim\sqrt{\hbar}$, this phase space scale being independent
of the value of $D$. However, the form of the large-scale classical
template is determined by the structures present at $t^*$, which is
sensitive to the value of $D$. Additionally, larger values of $D$ will
lead to stronger filtering in both the quantum and classical cases. At
a fixed value of time and $\hbar$, this means that slices of the
Wigner function at higher $D$ values will have broader features and
more efficient filtering of small scales. This aspect is clearly
demonstrated in the three panels of Figure~2.

\begin{figure}[htbp]
\includegraphics[width=3.6in]{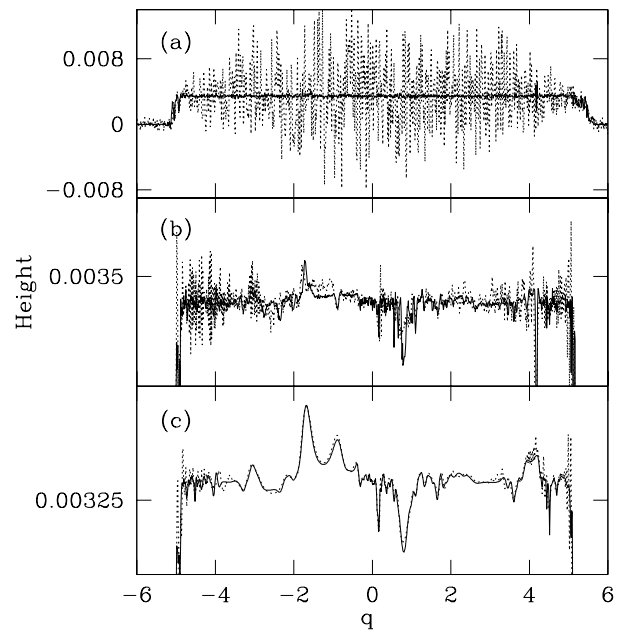}
\caption{\label{fig:SlicePlot} Sectional cuts of Wigner functions
  (dashed lines) and classical distributions (solid lines) for a
  driven Duffing oscillator, after 149 drive periods, taken at $p=0$
  for (a) $D=10^{-5}$; (b) $D=10^{-3}$; (c) $D=10^{-2}$. Parameter
  values are as stated in the text; the height is specified in scaled
  units.}
\end{figure}
We now examine the time-dependence in more detail. In Figure 3, we
display cross-sectional slices taken at $t=10$ and $t=30$.  At $t=10$,
the slice in the top panel, the classical and quantum functions have
still clearly not explored phase space sufficiently.  The Wigner
function, has significant negative values and the classical
distribution function has not been heavily broken up by the dynamics.
By $t=30$ (still less than $t_{qc}$), lower panel, the picture begins
to resemble the late-time plot shown in Figure 2. The negative regions
of the Wigner function have been largely eliminated and the functions
are distributed throughout phase space and are in approximate
agreement. This result is also consistent with the estimated value of
$t_{qc}$.
\begin{figure}[htbp]
\includegraphics[width=3.4in]{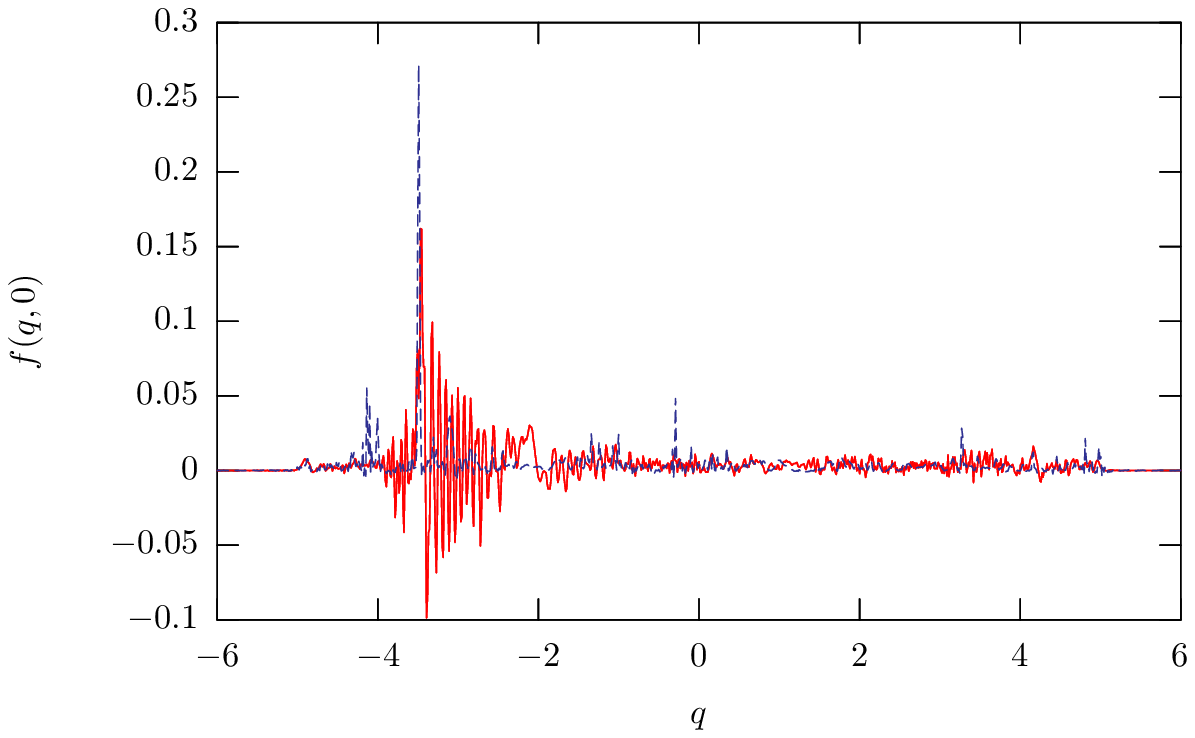}
\includegraphics[width=3.4in]{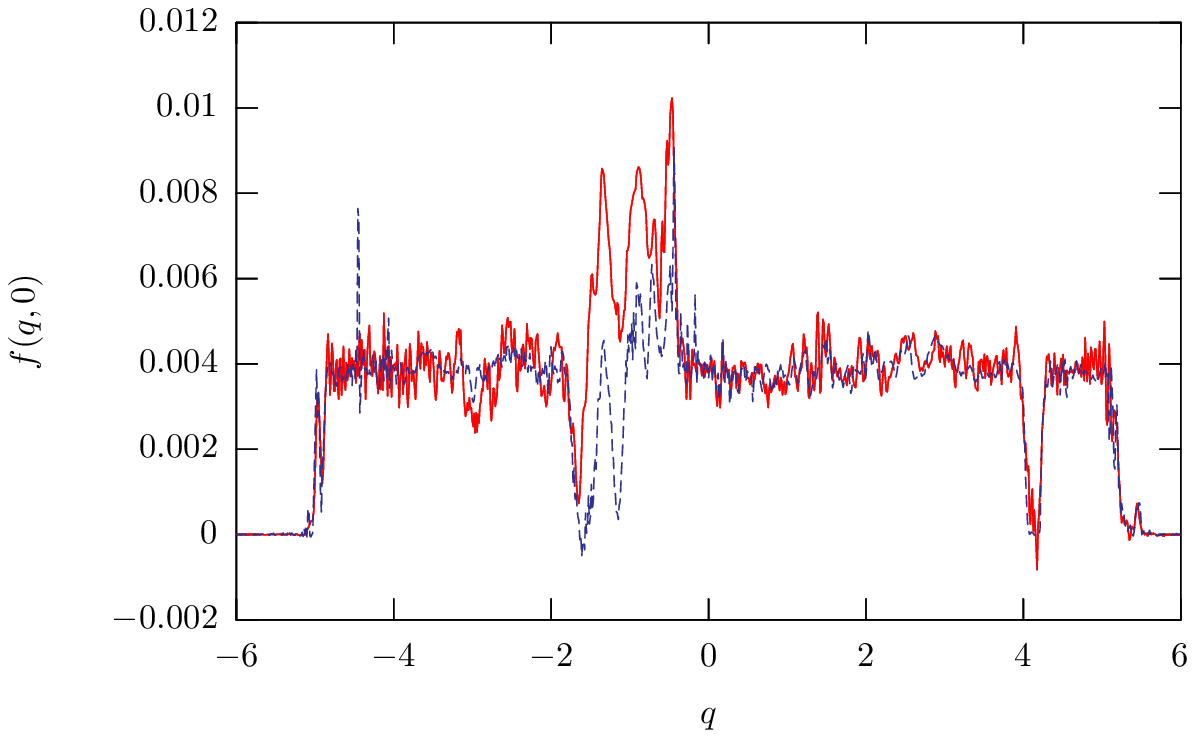}
\caption{\label{fig:h=0.1_10-3} Cross-sectional slices of the Wigner
  function (red) and classical distribution function (blue) taken in
  phase space for $p=0$ and $D=10^{-3}$.  The higher plot is taken at
  $t=10$ and the lower plot is taken at $t=30$.}
\end{figure}

We now perform a similar analysis for $D=10^{-2}$, for which $t_{qc}$
is an order of magnitude shorter. The top panel of Figure~4 is a
snapshot at $t=8$, close to $t_{qc}$, whereas the lower plot is taken
at the later time $t=20$. The early-time panel shows that the two
distributions closely agree on general features of the dynamics, as
predicted. By $t=20$, the weak QCT is well-stabilized, and one sees the
strong agreement on most individual features present in the late-time
case shown in Figure~2.
\begin{figure}[htbp]
\includegraphics[width=3.4in]{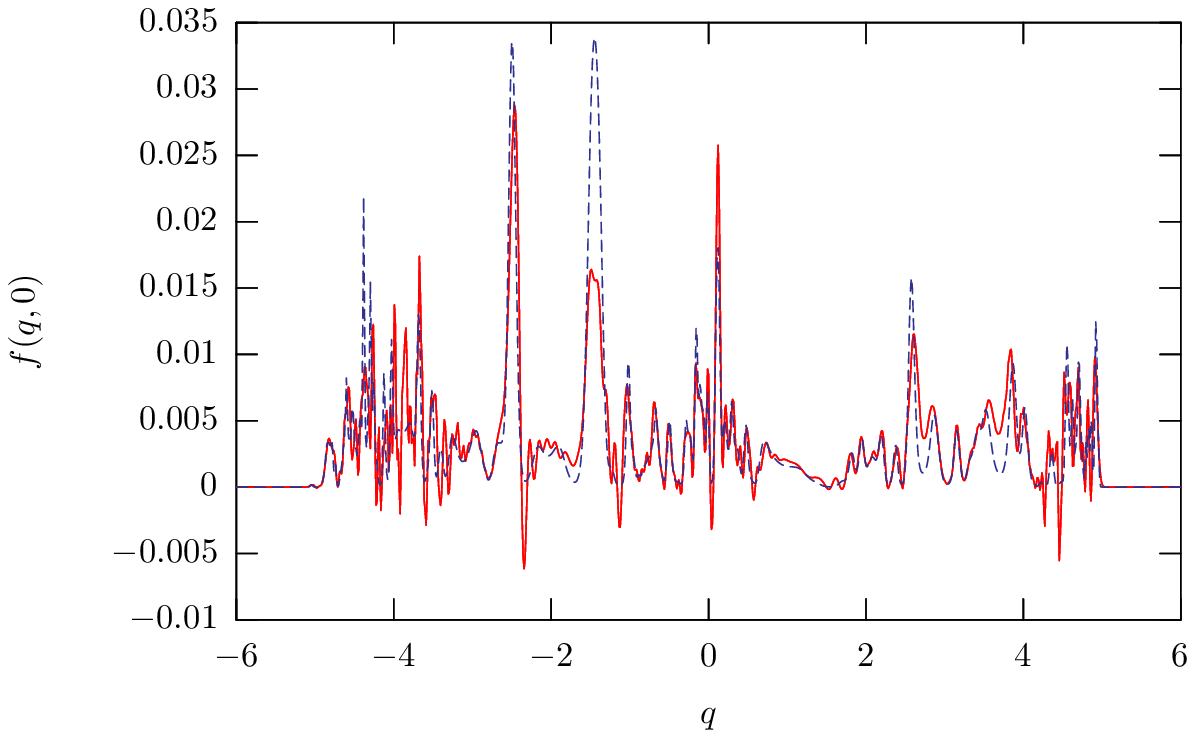}
\includegraphics[width=3.4in]{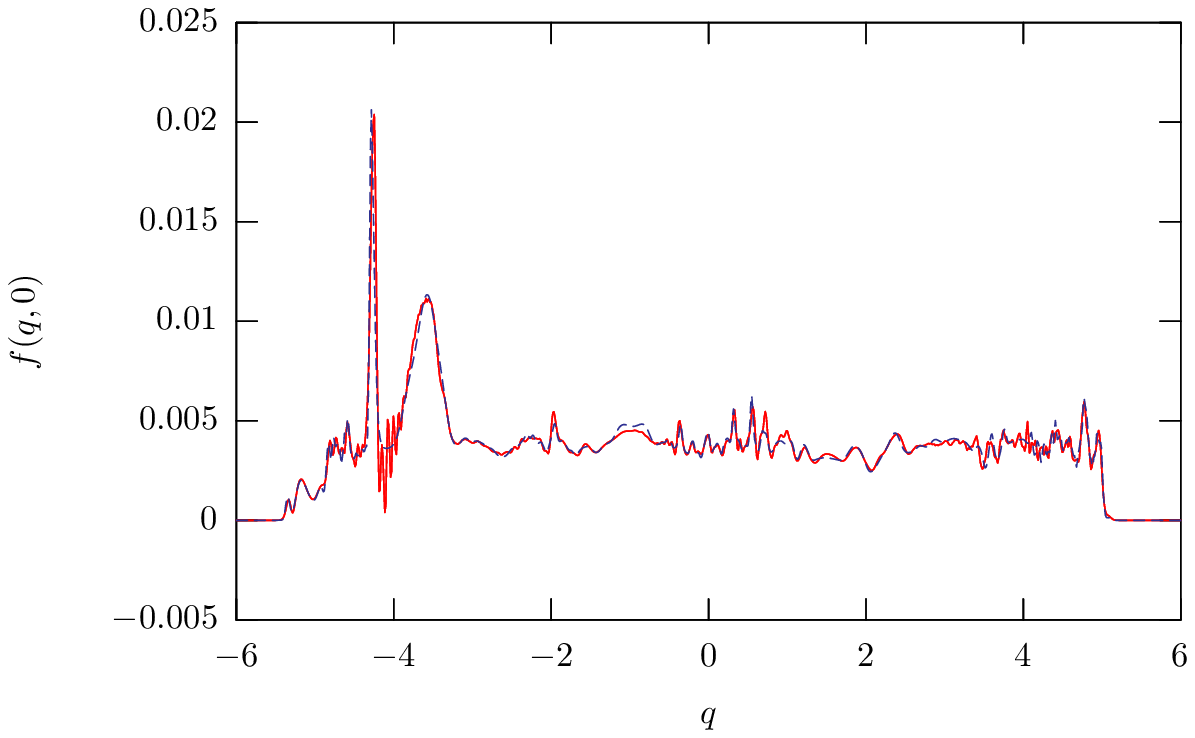}
\caption{\label{fig:h=0.1_10-2} Cross-sectional slices of the Wigner
  function (red) and classical distribution function (blue) taken in
  phase space for $p=0$ and $D=10^{-2}$.  The higher plot is taken at
  $t=8$ and the lower plot is taken at $t=20$.}
\end{figure}
As indicated earlier, similar results were seen at other values of
$\hbar$, at the same level of detail shown here for $\hbar=0.1$.  Many
of the studies of the late-time dynamics of this system appear in 
Ref.~\cite{BenPhD}. As a further example, we show plots in Figure~5 
for the case of $\hbar=1$ and with stronger
noise coupling than the typical case considered in the theoretical
analysis.  In the upper panel $D=0.1$, while the lower plot has $D=1$;
the snapshots are taken at $t=20$. (In both cases, $t_{qc}<t^*$.) The
top slice shows general agreement between the two distributions, with
some residual quantum interference effects. When $D=1$, the
formal value of $t_{qc}$ is less than the dynamical timescale. This
indicates that the onset of the weak QCT should be very rapid and by
the relatively late time at which the snapshot is taken, the
transition should be complete. The numerical results are very
consistent with this prediction. These two plots show that even in the
regime where our formal analysis might break down (large $\hbar$,
large $D$), the general features and timescales follow the predicted
estimates. At even larger (unphysical) values of $\hbar$ and $D$, the
phase space boundaries become important and the smoothing gets so
large that dynamical features hardly survive in the distributions.

\begin{figure}[htbp]
\includegraphics[width=3.4in]{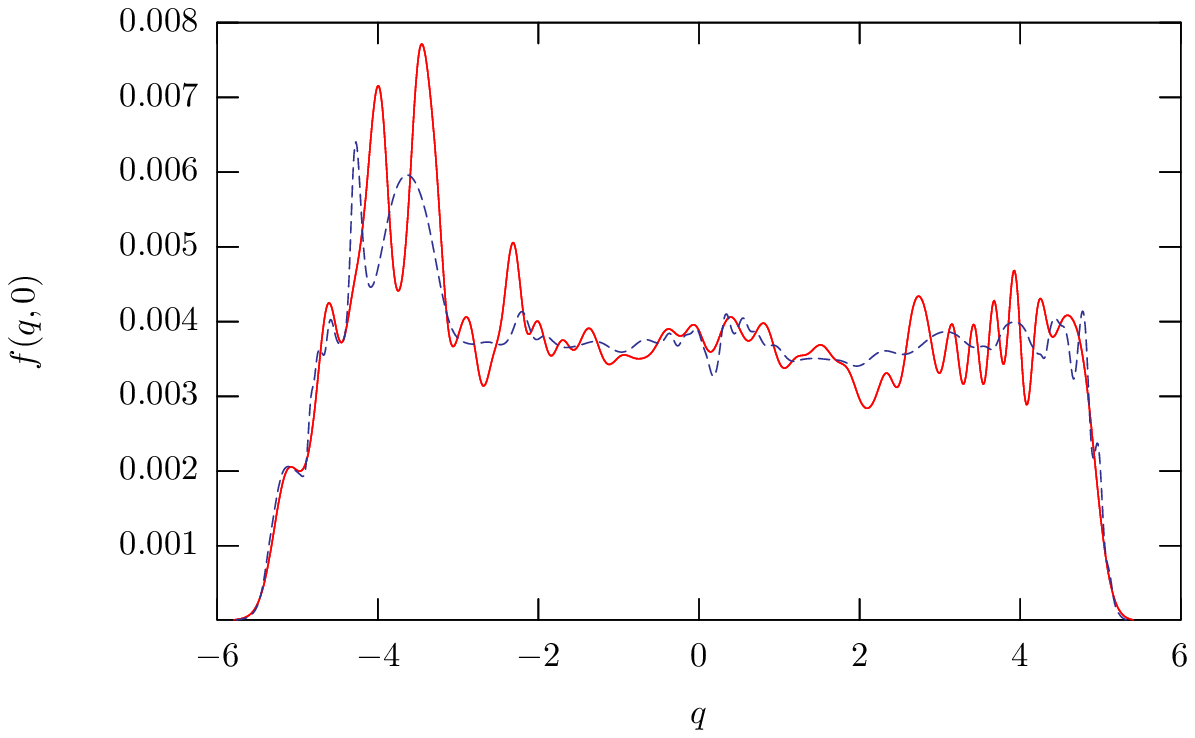}
\includegraphics[width=3.4in]{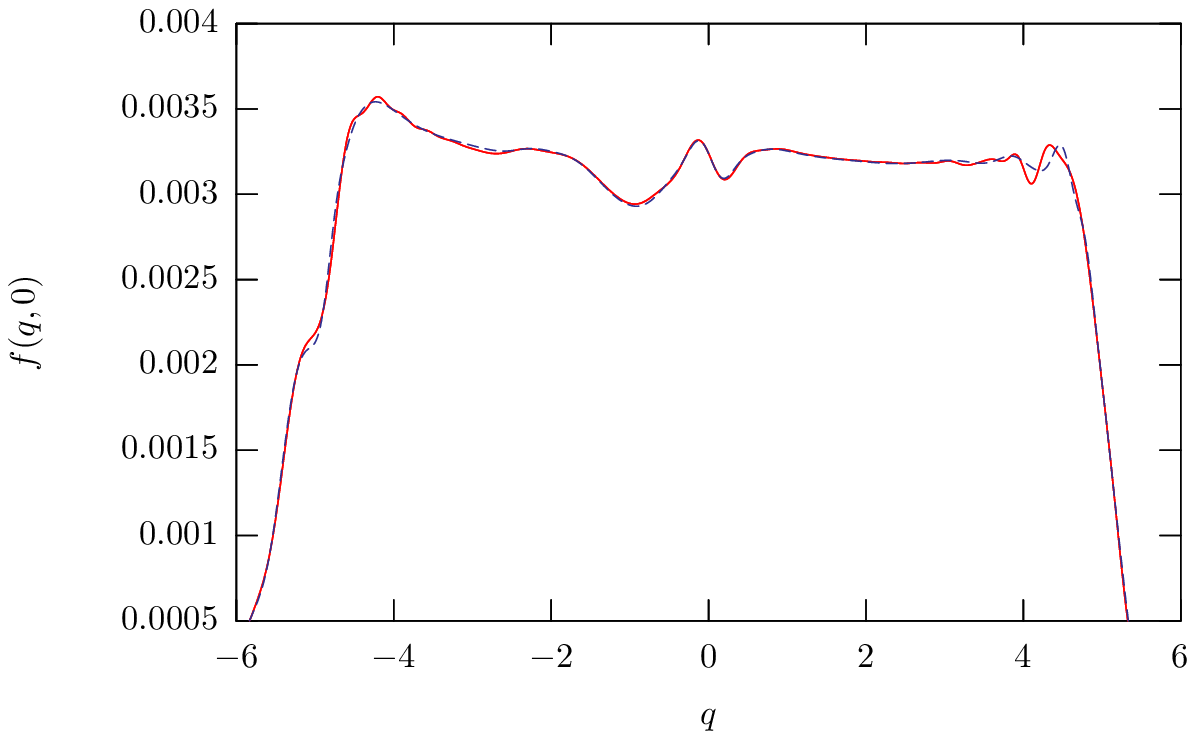}
\caption{\label{fig:h=1_020} Cross-sectional slices of the Wigner
  function (red) and classical distribution function (blue) following
  Figure~4.  The upper panel has $D=0.1$ and the lower panel, $D=1$.
  Both snapshots are taken for at $t=20$.}
\end{figure}

Finally, we address one last point of the argument -- that the
noise-averaged termination of fine scale structure would lead to the
presence of the early time folding associated with the foliation of
the unstable manifold generated by the homoclinic point of the Duffing
system. Evidence for this is presented in Figure~7, a full late-time,
high-resolution phase space rendering of the Wigner function for
$D=10^{-3}$.  The time is taken to be $t=149$ (roughly a factor of two
greater than $t_{qc}$), well after the quantum-classical transition
has occurred.  Superimposed on the Wigner function is the early time
unstable manifold. It is clear that the evolution has organized along
these early-time features, as expected from our analysis. The final
distribution which both the dual classical distribution and Wigner
function approach, once the transition has occurred, shows the
suppression of the late-time, fine-scale features of the unstable
manifold, as it is supported by the large early-time structures.
Quantum interference, while expected, is local and is strongest near
the sharp turns in the manifold where branches are most close
together.  This, combined with the previous results in this section,
allow us to conclude that the basic mechanisms posited for the
quantum-classical transition are consistent with results from
numerical simulations.
\begin{figure}[htbp]
\begin{flushleft}
\includegraphics[width=3.6in]{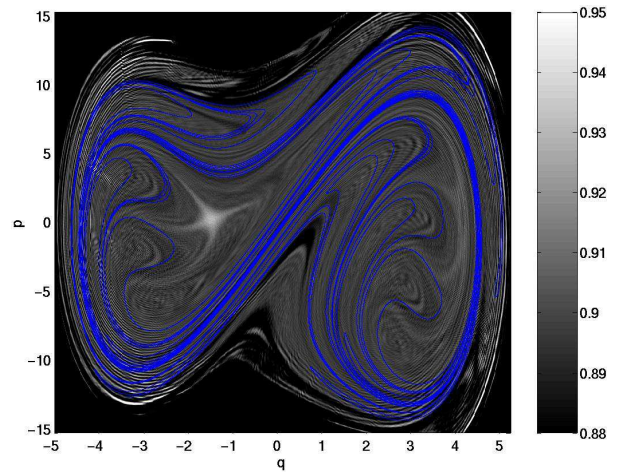}
\end{flushleft}
\caption{\label{fig:WigMan} Phase space rendering of the Wigner
  function at time $t=149$ periods of driving. The early-time part of
  the unstable manifold associated with the noise-free dynamics is
  shown in blue (see text for discussion). The value of $D=10^{-3}$ is
  not sufficient to wipe out all the quantum interference which, as
  expected, is most prominent near sharp turns in the manifold.}
\end{figure}

\section{Conclusion and Future Directions}

We have presented a set of physical mechanisms which explain the
source of the weak quantum-classical transition for one-dimensional,
bounded chaotic systems. The fact that one-dimensional, chaotic
systems are being investigated in real world laboratory experiments,
with interesting potential applications, further enhances the
importance of understanding how unconditioned environmental
interactions affect a subsystem of interest. We have used this
understanding to derive estimates for the time at which the weak
transition occurs.

It is important to keep in mind that currently there is no general
understanding of which systems will actually exhibit a weak QCT, so
the existence of this timescale has another useful feature, in terms
of classification of quantum dynamical systems. If a weak QCT has not
occurred by the predicted $t_{qc}$, our analysis would argue that it
will not occur at all (within the parametric assumptions made). So the
existence or nonexistence of this time can be used as a test for the
occurrence of long-time quantum-classical correspondence (but still on
timescales shorter than the physical equilibration timescale) without
any knowledge of initial conditions.

Our numerical results illustrated the compact manifold structure
induced by the bounded phase space region.  The role of boundedness is
a key component in the theoretical analysis presented earlier in this
paper.  This topological feature causes the system to fold on itself,
which, in turn, allowed us to estimate a timescale for the termination
of fine structure.  A one-dimensional bounded chaotic evolution,
coupled with noise, appears to necessarily terminate fine scale
structure.  In order for the analysis to be valid, the system must be
bounded or, if it is unbounded, it must at least fold onto itself in
such a way as to allow a similar process to take place.  The lack of
such an evolution may be a reason why no such transition was found for
the manifestly unbounded delta-kicked rotor studied in
Ref.~\cite{bhjs}. Additionally, while an unconditioned evolution
eliminated all negativity from the quantum Wigner function, the
distribution was still not classical~\cite{hjmrss}.

It is useful to restate the ways in which the present analysis differs
from previous work. First, the connection to continuous measurement
and the weak and strong forms of quantum-classical correspondence are
explicitly stated. Second, we use the dual classical Fokker-Planck
equation not to represent a physical classical evolution, but rather
as a dynamical foil of the open-system quantum evolution, one that can
handle quantum initial states, and quantum backaction (which is
missing in classical theory), but keeps only the classical system
propagator. Third, our analysis is symmetric -- we consider the effect
of noise acting as a filter on the open-system quantum evolution
(treated in semiclassical approximation) melded with a consideration
of noise-induced filtering on the classical dual evolution with its
exponential-in-time folding of phase space structures characteristic
of chaos. This folding points to the role of global phase-space
topology in deriving our results, and distinguishes them from local,
heuristic analyses of the role of decoherence in the quantum to
classical transition~\cite{ZP}.

Clearly, more work is needed to fully explore the conditions under
which the weak QCT exists, especially the role of boundedness.  In
this regard, investigation of two-dimensional systems would be
informative as several of the topological arguments presented here
would likely need to be modified. Adding more dimensions would
introduce effects such as Arnold diffusion which become important
components of the dynamics.  Many of these features lack a lower
dimensional analog, so it is reasonable to believe that they could
play an important role in the higher dimensional QCT.  It would also
be interesting to see what qualitative features of the weak quantum to
classical transition will be preserved in these systems.  More
immediately, the connection between the requirements for weak and
strong QCT scenarios are worth contrasting, especially to delineate
parameters regimes for validity.  This project is currently
underway~\cite{bkbk}.
 
Numerical simulations were performed on the Cray T3E and IBM SP3 at
NERSC, LBNL. B.D.G. and S.H. acknowledge support from the LDRD program
at LANL. B.D.G. was also partially supported by the Center for
Nonlinear Studies at LANL.  The work of B.S. and B.D.G (partially) was
supported by the National Science Foundation grant \#0099431.

\end{document}